\documentclass[%
 reprint,
groupedaddress,
frontmatterverbose, 
 amsmath,amssymb,
 aps,
 showkeys,
 showpacs,
pra,
]{revtex4-2}
\usepackage{dcolumn}
\usepackage{bm}
\usepackage{amsmath}
\usepackage{hyperref}       
\usepackage{url}            
\usepackage{booktabs}       
\usepackage{amsfonts}       
\usepackage{xcolor}         
\usepackage[thinlines]{easytable}
\usepackage{subfig}
\usepackage{microtype}
\usepackage{nicefrac}       
\usepackage[export]{adjustbox}
\usepackage{caption}
\usepackage{nccmath}
\usepackage{graphicx}
\usepackage{float}
\newcommand{\vect}[1]{\boldsymbol{#1}}

\hypersetup{
    colorlinks,
    linkcolor={red!50!black},
    citecolor={blue!80!black},
    urlcolor={blue!80!black}
}
\usepackage{titlesec}
\titleformat{\subsection}[hang]{\normalfont\small\bfseries}{\thesubsection}{1em}{}
\titleformat{\subsubsection}[hang]{\normalfont\footnotesize\bfseries}{\thesubsubsection}{1em}{}
\titlespacing{\section}{0pt}{*1.5}{*1}
\titlespacing{\subsection}{0pt}{*1.5}{*1}
\titlespacing{\subsubsection}{0pt}{*1.5}{*1}
\usepackage{setspace}
\usepackage[most]{tcolorbox}

\begin{document}
\title{High-Order Triadic Functional Connectivity in the Brain and Beyond}
\author{Qiang Li$^{1}$}
\email{Corresponding author: qli27@gsu.edu}
\author{Masoud Seraji$^{1,2}$}
\author{Yu-Ping Wang$^{3}$}
\author{Godfrey D Pearlson$^{4,5,6}$}
\author{Vince D Calhoun$^{1,7,8,9}$}
\affiliation{
$^{1}$Tri-Institutional Center for Translational Research in Neuroimaging and Data Science (TReNDS), Georgia State, Georgia Tech, and Emory University, Atlanta, GA, United States}
\affiliation{$^{2}$Department of Psychology, University of Texas at Austin, Austin, TX, United States}
\affiliation{$^{3}$Department of Biomedical Engineering, Tulane University, New Orleans, LA, United States}
\affiliation{$^{4}$Department of Psychiatry, Yale University, New Haven, CT, United States}
\affiliation{$^{5}$Department of Neuroscience, Yale University, New Haven, CT, United States}
\affiliation{$^{6}$Olin Neuropsychiatry Research Center, Hartford Hospital, Hartford, CT, United States}
\affiliation{$^{7}$Neuroscience Institute, Georgia State University, Atlanta, GA, United States}
\affiliation{$^{8}$Psychology Department, Georgia State University, Atlanta, GA, United States}
\affiliation{$^{9}$Department of Computer Science, Georgia State University, Atlanta, GA, United States} 

\begin{abstract}
Here, we report high-order functional network connectivity as a promising way for studying the brain connectome. Traditional functional connectivity approaches capture only pairwise relationships between brain regions, overlooking complex multivariate dependencies that underlie cognition and behavior. First, we demonstrated that high-order interactions capture more information and can distinguish between resting-state and task-state brain activity. Second, we introduce a matrix-based entropy-functional method for estimating triadic interactions, which are statistical dependencies among triplets of brain regions, and apply it to large-scale functional brain networks. The resulting triadic networks revealed distinct community patterns that complement those observed in traditional pairwise functional connectivity analyses and simultaneously capture additional connection information. Despite the potential combinatorial explosion of triadic configurations, the networks exhibited constrained and hierarchical structures that allowed computation and interpretation. These findings position triadic connectivity as a promising next-step functional connectivity framework for probing brain network organization and high-order neural interactions, while also highlighting key biological and technical challenges that require careful consideration. \\

\noindent \textbf{Functional connectivity is widely used in brain imaging but typically captures only second-order, pairwise interactions. Large-scale brain networks, however, exhibit high-order interactions that go beyond such analyses. Here, we introduce a high-order triadic functional connectivity matrix, enabling the characterization of three-way interactions in fMRI brain signals. We highlight emerging opportunities and discuss both biological and methodological challenges. High-order triadic functional connectivity provides a novel, informative framework for advancing brain imaging and multimodal signal analysis beyond traditional pairwise measures.}
\end{abstract}
\maketitle

\section*{Introduction}
Brain function arises from a web of nonlinear connections, with activity that spreads unevenly across both space and time~\cite{Vince14neuron,Li25bp}. Consequently, a quantitative description of cognition, perception, and behavior requires a precise characterization of the functional couplings between its constituent regions. Functional connectivity, which quantifies statistical dependencies between brain signals, provides a principled framework to probe interactions between brain regions~\cite{Sporns05}. This approach has revealed the large-scale architecture of functional brain networks, their modular organization, and their state-dependent reconfigurations, establishing functional connectivity, and its network-level counterpart, functional network connectivity, as a fundamental framework for studying both healthy brain dynamics and network disruptions in neurological and psychiatric disorders~\cite{Rogers08,Woodward15,Qiang25mp}.

Most functional connectivity analyses to date have focused on pairwise interactions, characterizing second-order statistical dependencies between regional activity signals. While pairwise approaches have provided valuable insights, they inherently overlook high-order interactions among multiple regions that may underlie coordinated neural computation~\cite{Battiston20}. Such interactions can reflect cooperative or competitive dynamics that pairwise analyses cannot detect, limiting our understanding of the brain’s multivariate information flow and its complex functional architecture. 

Recent advances in theory and computation now allow us to move beyond pairwise functional connectivity and quantify triadic and high-order interactions among brain regions~\cite{Varley23pnas,QiangNC23,Lihbp25}. These measures capture combinatorial dependencies, revealing network structures hidden from conventional analyses. Integrating high-order connectivity into network models provides a more realistic account of how information is distributed and integrated across the brain. This study presents a high-order triadic functional connectivity framework, offering a new approach for exploring brain network organization beyond traditional pairwise methods. By capturing interactions among triplets of regions, this approach provides a more comprehensive view of functional connectivity and opens new avenues for understanding how distributed neural activity supports cognition and behavior.

\section*{Materials and Methods}
\textbf{Resting-State and Task-Based fMRI Datasets.} The resting-state fMRI (rsfMRI) data included 600 unrelated healthy subjects, all recruited from the multi-site Bipolar and Schizophrenia Network on Intermediate Phenotypes study~\cite{tamminga2013clinical,Shashwath14pnas}. All participants were psychiatrically stable and maintained on stable medication regimens at the time of the study. During scanning, they were instructed to rest with their eyes closed while remaining awake. Detailed acquisition and scanning parameters for the full study sample are described elsewhere~\cite{tamminga2013clinical,Shashwath14pnas}. For the task fMRI dataset, we analyzed scans from 66 participants selected from the publicly available \textit{Nilearn} brain developmental dataset~\cite{Hilary18nc}, including 33 children (ages 3–7) and 33 adults (ages 18–39). During scanning, participants viewed the Disney Pixar short film Partly Cloudy without performing any additional tasks. The movie commenced after a brief rest period (black screen; TRs 0–5), and the initial segment corresponding to the opening credits lasted from TRs 6–10.

\textbf{rsfMRI and Task fMRI Analyses.} rsfMRI and task fMRI data were preprocessed using FSL 5.0.11 and SPM12. The preprocessing pipeline included distortion correction, realignment to correct for head motion, normalization to MNI space, resampling to 3mm isotropic voxels, and spatial smoothing with a 6mm full width at half maximum (FWHM) Gaussian kernel. After preprocessing, spatially constrained independent component analysis (scICA) was applied using the NeuroMark\_fMRI\_2.2 template~\cite{Iraji2022CanonicalAR} as a prior to extract subject-specific time series and estimate functional connectivity matrices. This procedure yielded 105 intrinsic connectivity networks (ICNs) spanning multiple spatial resolutions (\textbf{\textit{Fig.\ref{fig:0}A}}). These networks were derived from over 20 studies using a group multi-scale ICA approach~\cite{Iraji2022CanonicalAR}, encompassing eight distinct model orders (e.g., 25, 50, 75, 100, 125, 150, 175, and 200) to capture functional networks at varying spatial scales. Higher model orders correspond to finer spatial resolution, whereas lower orders integrate features across broader brain regions. This multi-scale framework allows for comprehensive modeling of diverse ICNs. Of the 900 components initially extracted, 105 were retained after manual selection based on criteria including peak activation in gray matter, minimal overlap with vascular or ventricular structures, and low similarity to artifacts.

\textbf{Multivariate Objective Optimization ICA.} Multivariate Objective Optimization ICA with Reference (MOO-ICAR) is a spatially constrained ICA (scICA) method developed to extract subject-specific independent components (ICs) by leveraging predefined network templates as spatial priors~\cite{Yuhui20,Calhoun11FronPsy}. This approach improves the consistency of network identification across individuals by aligning component estimation with established functional network structures. A key advantage of the MOO-ICAR framework is its ability to preserve inter-subject correspondence during component estimation. Moreover, the scICA framework allows flexible template selection, enabling researchers to tailor analyses for condition-specific studies or broader investigations of canonical brain networks across diverse populations~\cite{Calhoun2001AMF}.

The MOO-ICAR algorithm simultaneously optimizes two objective functions to extract subject-specific ICs: one objective maximizes the statistical independence of the estimated networks, while the other ensures spatial similarity with a predefined network template~\cite{Yuhui20,Iraji2022CanonicalAR}. The following equation illustrates how the $l^{\text{th}}$ network component for the $k^{\text{th}}$ subject is estimated using the corresponding template $S_l$ as a spatial reference:

\begin{equation}
\begin{split}
& \max \left\{
\begin{array}{l}
J\left(S_l^k\right) = \left\{ E\left[G\left(S_l^k\right)\right] - E[G(v)] \right\}^2 \\
F\left(S_l^k\right) = E\left[ S_l S_l^k \right],
\end{array}
\right. \quad \\
& \text{s.t.} \ \left\|w_l^k\right\| = 1.
\end{split}
\end{equation}

Here, $S_l^k = \left(w_l^k\right)^T X^k$ denotes the estimated $l^{\text{th}}$ network for the $k^{\text{th}}$ subject, where $X^k$ is the whitened fMRI data matrix and $w_l^k$ is the associated unmixing column vector to be optimized. The term $J(S_l^k)$ quantifies independence through a measure of negentropy, where $G(\cdot)$ is a non-quadratic function, $v$ is a standard Gaussian variable (mean zero, unit variance), and $E[\cdot]$ represents expectation. The term $F(S_l^k)$ quantifies the similarity between the subject-specific component and the network template. These two objectives are combined into a single cost function using a linear weighted sum, with equal weights of 0.5. Applying MOO-ICAR within the scICA framework yields individualized ICNs for each of the $N=105$ predefined templates along with their associated time courses, ensuring consistent network alignment across subjects.

\textbf{High-Order Triadic Functional Network Connectivity.} We estimated interactions beyond pairwise (i.e., triadic, $\vect{k} = 3$) among $\vect{n} = 105$ ICNs by iterating over all sets of indices to compute the total correlation  (\emph{TC})~\cite{watanabe1960information} (\textbf{\textit{Fig.\ref{fig:0}B}}). Strong interactions among ICNs result in higher TC values, whereas weaker interactions produce lower values. Notably, TC is always non-negative.

\subsection*{\emph{Renyi’s $\alpha$ Entropy} Functional}
In information theory, \emph{Rényi's $\alpha$-entropy}~\cite{renyi1961measures} serves as a natural generalization of the classical \emph{Shannon entropy}~\cite{shannon1948mathematical}. For a random variable $X$ with probability density function $p(x)$ in a finite set $\mathcal{X}$, the $\alpha$ entropy is defined as:
\begin{equation}
    \mathbf{H}_\alpha(X)=\frac{1}{1-\alpha} \log \left(\int_{\mathcal{X}} p^{\alpha}(x) dx \right),
\end{equation}
with $\alpha \neq 1$ and $\alpha \ge 0$. In the limiting case where $\alpha \rightarrow 1$, it reduces to \emph{Shannon’s entropy}~\cite{cover1991information}. 

\subsection*{Matrix-Based Information Estimator}
In practice, given $m$ realizations sampled from $p(x)$, i.e., $\{x_i\}_{i=1}^m$, Sanchez~Giraldo \emph{et al.}~\cite{giraldo2014measures} suggests that one can evaluate $\mathbf{H}_\alpha(X)$ without estimating $p(x)$. Specifically, the so-called \emph{matrix-based Renyi’s $\alpha$ entropy} is given as follows:
\begin{equation}
    \mathbf{H}_\alpha(X)=\frac{1}{1-\alpha} \log \left(\operatorname{tr}\left(A^\alpha\right)\right),
    \label{eq.ep}
\end{equation}
where $A \in \mathbb{R}^{m \times m}$ is a (normalized) Gram matrix with elements $A_{i j} = K_{i j}/\operatorname{tr}(K)$, $K_{i j}=\kappa\left(x_i, x_j\right)$ in which $\kappa$ stands for a positive definite and infinitely divisible kernel such as Gaussian. $\operatorname{tr}(*)$ refers to matrix trace. As in~\cite{yu2019multivariate}, \textcolor{black}{We set $\alpha=1.01$ specifically because it closely approximates Shannon entropy while providing numerical stability, and we chose a Gaussian kernel $G_\sigma$ with width $\sigma$, given by:}

\begin{equation}
    G_\sigma\left(x_i, x_j\right)=\beta \exp \left(-\frac{\left\|x_i-x_j\right\|^2}{2 \sigma^2}\right),
    \label{eq.Gaukernel}
\end{equation}
where $\beta$ is a constant whose value is irrelevant because it is canceled out in the normalized Gram matrix.

\subsection*{Estimating High-Order Triadic Dependencies through \emph{Matrix-based Rényi's $\alpha$} \emph{TC} and \emph{DTC}}
Suppose now we have $n\geq2$ variables ($X^1,X^2,\cdots,X^n$) and a collection of $m$ samples (Throughout this paper, we use superscript to denote variable index and subscript to denote sample index. For example, $x^3_i$ refers to the $i$-th sample from the $3$rd variable.), i.e., $\{x^1_i,x^2_i,\cdots,x^n_i\}_{i=1}^m$, the \emph{matrix-based Renyi’s $\alpha$ joint entropy} for $n$ variables can be evaluated as~\cite{yu2019multivariate}: 

\begin{equation}
\small
\begin{split}    
    & \mathbf{H}_\alpha\left(X^1, X^2, \cdots, X^n\right)=\mathbf{H}_\alpha\left(\frac{K^1 \odot K^2 \odot \cdots \odot K^n}{\operatorname{tr}\left(K^1 \odot K^2 \odot \cdots \odot K^n\right)}\right) \\ 
    & = \frac{1}{1-\alpha} \log \left(\operatorname{tr}\left(\left(\frac{K^1 \odot K^2 \odot \cdots \odot K^n}{\operatorname{tr}(K^1 \odot K^2 \odot \cdots \odot K^n)}\right)^\alpha\right)\right),
    \label{eq.muljoin}
\end{split}
\end{equation}

where $\left(K^n\right)_{i j} = \kappa(x^n_i,x^n_j) \in \mathbb{R}^{m \times m}$ is a Gram matrix evaluated with $\kappa$ for the $n$-th variable. The operator $\odot$ is the Hadamard product.

\emph{TC} quantifies the dependence among $n$ variables and can be viewed as a non-negative extension of \emph{mutual information} from two variables to $n$ variables. Following Watanabe~\cite{watanabe1960information}, the \emph{TC} is defined as:

\begin{equation}\label{eq.tc}
\small
\begin{split}
   & \mathbf{TC}\left(X^1, \cdots, X^{n}\right) = \sum_{i=1}^n \mathbf{H}\left(X^{i}\right)-\mathbf{H}\left(X^1, \cdots, X^{n}\right).
\end{split}
\end{equation}

In order to estimate \emph{TC} in a practical setting only from $m$ samples $\{x^1_i,x^2_i,\cdots,x^n_i\}_{i=1}^m$, we convert (\ref{eq.tc}) to matrix-based \emph{Renyi's $\alpha$ entropy functional} based on (\ref{eq.ep}) and (\ref{eq.muljoin}), which simplifies the estimation~\cite{yu2019multivariate} and enables a reformulation:

\begin{equation}
\small
\begin{split}
& \mathbf{TC}_\alpha(X^1,X^2,\cdots,X^n)=\sum_{i=1}^n \mathbf{H}_\alpha(X^i) -\mathbf{H}_\alpha(X^1,X^2,\cdots,X^n) \\
& =\left[ \sum_{i=1}^n \frac{1}{1-\alpha} \log \left(\operatorname{tr}\left(\frac{K^i}{\operatorname{tr}(K^i)}\right)^\alpha\right) \right] - \\ & \frac{1}{1-\alpha} \log \left(\operatorname{tr}\left(\left(\frac{K^1 \odot K^2 \odot \cdots \odot K^n}{\operatorname{tr}(K^1 \odot K^2 \odot \cdots \odot K^n)}\right)^\alpha\right)\right).
\end{split}
\label{eq.tca}
\end{equation}

The dual total correlation (\emph{DTC})~\cite{Han78}, also known as binding information, provides a complementary perspective by emphasizing shared dependencies. Mathematically, \emph{DTC} is defined as
\begin{equation}
    \begin{split}
    & \textbf{DTC}_\alpha(X^1, \dots, X^n) = \\ 
    & \left[\sum_{i=1}^n \mathbf{H}_\alpha\left(X^{[n] \backslash i}\right)\right]-(n-1) \mathbf{H}_\alpha\left(X^1, X^2, \cdots, X^n\right).
\end{split}
\end{equation}

\emph{DTC} provides an alternative decomposition of multivariate information, highlighting interactions not fully captured by \emph{TC} alone. 

\begin{figure*}[!ht]
    \centering
    \includegraphics[width=0.95\textwidth, height=11cm]{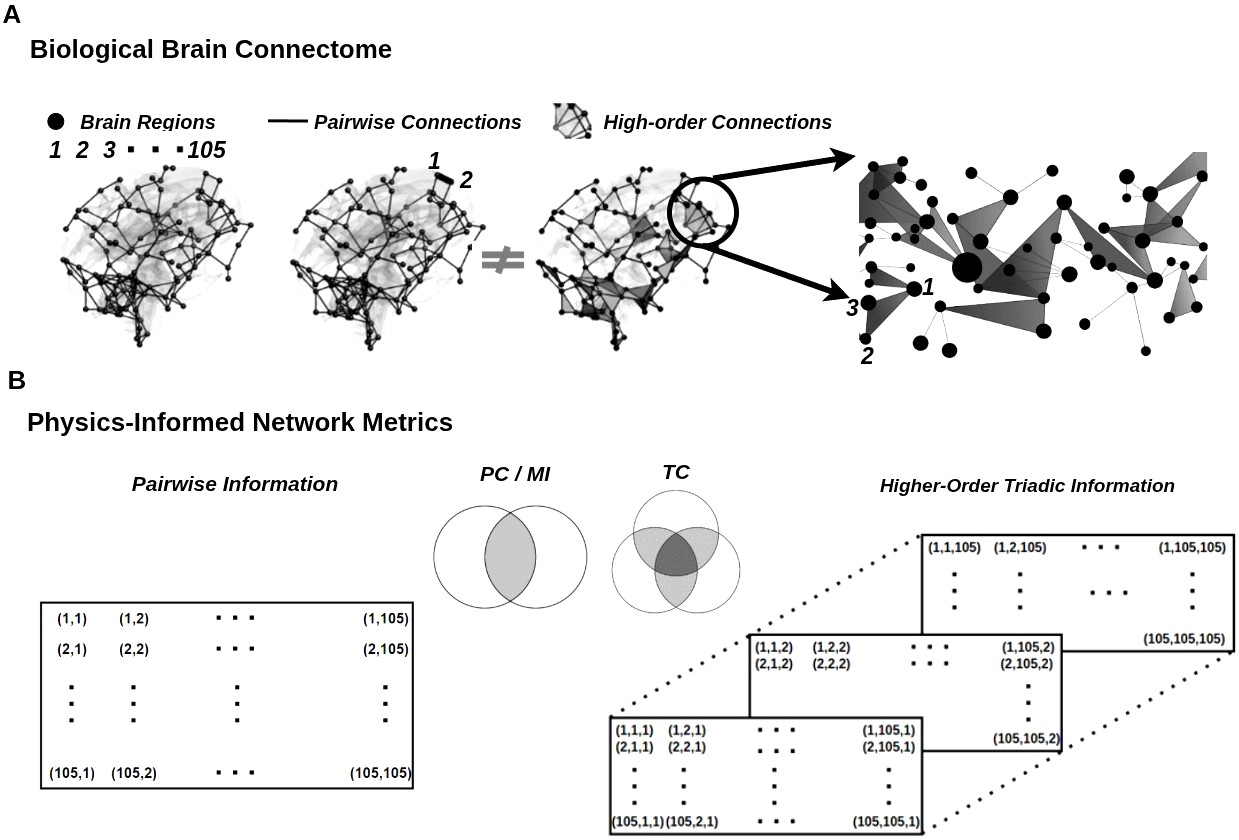}
    \caption{Brain connectivity modes. Brain regions are represented as a network, where interactions occur through both pairwise and high-order connections (\textbf{A}). Pairwise connections alone are insufficient to capture complex interactions. To quantify these interactions, physics-informed network metrics are employed: Pearson correlation (PC) and mutual information (MI) for pairwise connections, and total correlation (TC), an extension of MI, for high-order interactions in the brain (\textbf{B}).}
    \label{fig:0}
\end{figure*}

\textbf{Pairwise Functional Connectivity.} The resting-state fMRI signals consist of $n$ ICNs derived from scICA, denoted as $X_i$ ($1 \le i \le n$, with $n=105$). Each $X_i$ represents the signal at a given time point $t_i$ within the sequence $\{t_1, t_2, \ldots, t_T\}$, sampled at a constant interval $t$. Pairwise interactions between ICNs can be assessed using Pearson correlation, where ICNs sharing substantial information are expected to exhibit strong correlations, while weakly related ICNs show lower correlations (\textbf{\textit{Fig.\ref{fig:0}B}}). Alternatively, pairwise dependencies can also be quantified using mutual information. Mutual information captures the dependency between $X_i(t)$ and $X_j(t)$, and attains a minimum value of zero when the two ICNs are statistically independent.

\section*{Results}
\subsection*{High-Order Interaction Patterns Differentiate Information Structure Between Resting-State and Task Conditions}
\begin{figure*}[htbp]
    \centering
    \includegraphics[width=0.6\textwidth, height=4.3cm]{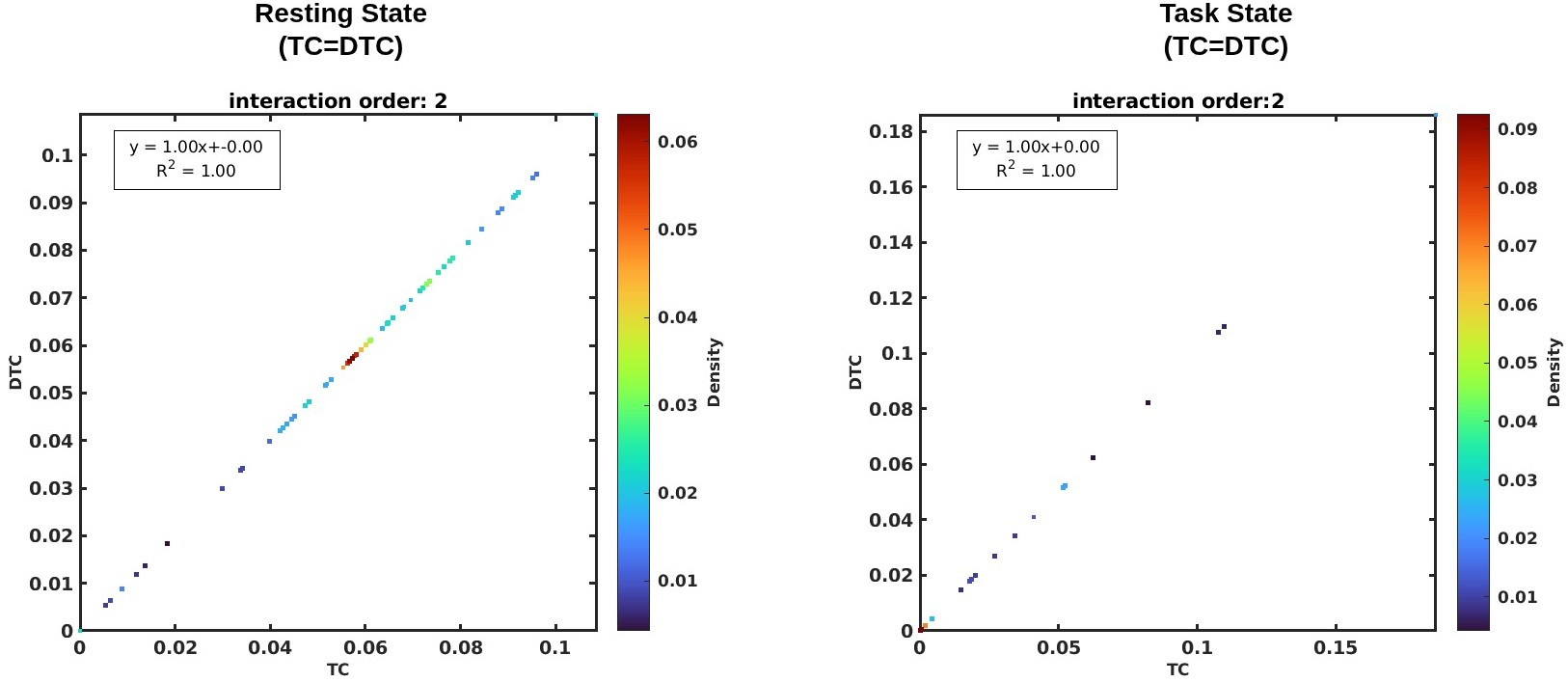} 
    \rule{\textwidth}{0.5mm}
    \includegraphics[width=\textwidth, height=7.2cm]{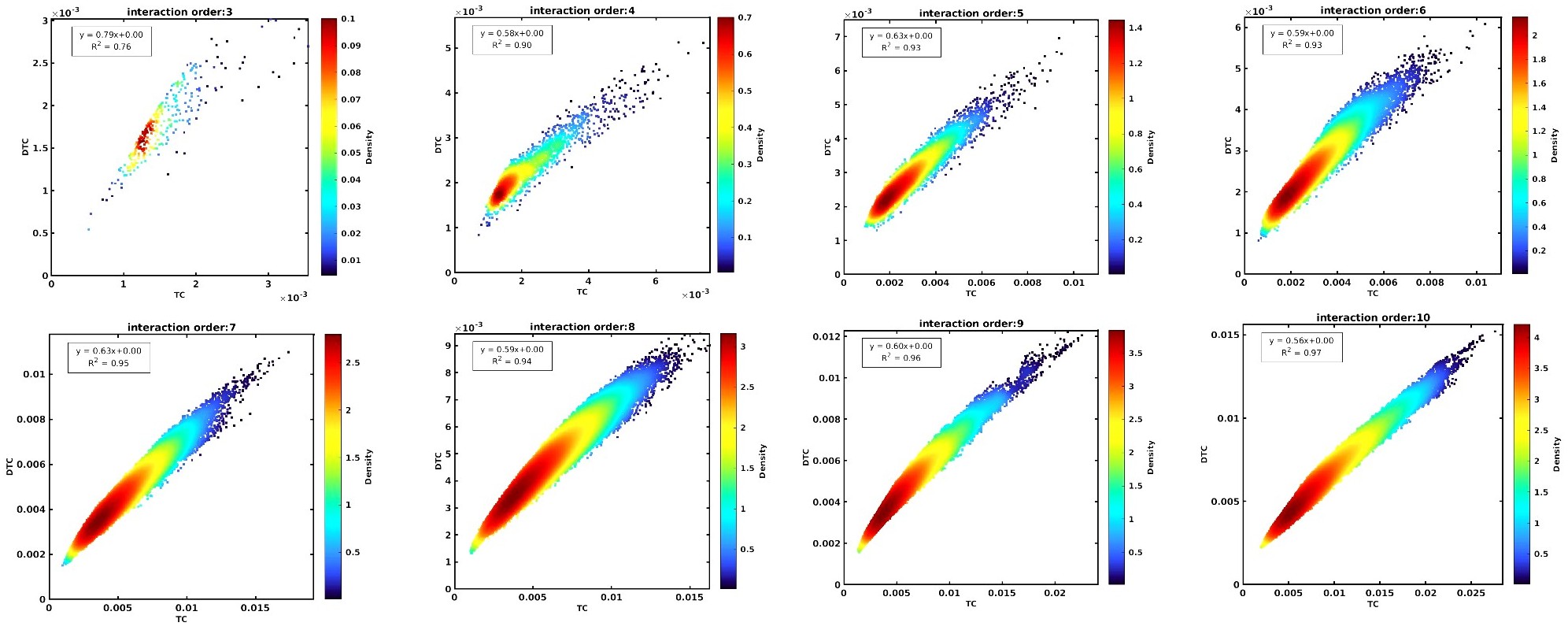} 
    \rule{\textwidth}{0.5mm}
    \includegraphics[width=\textwidth, height=7.2cm]{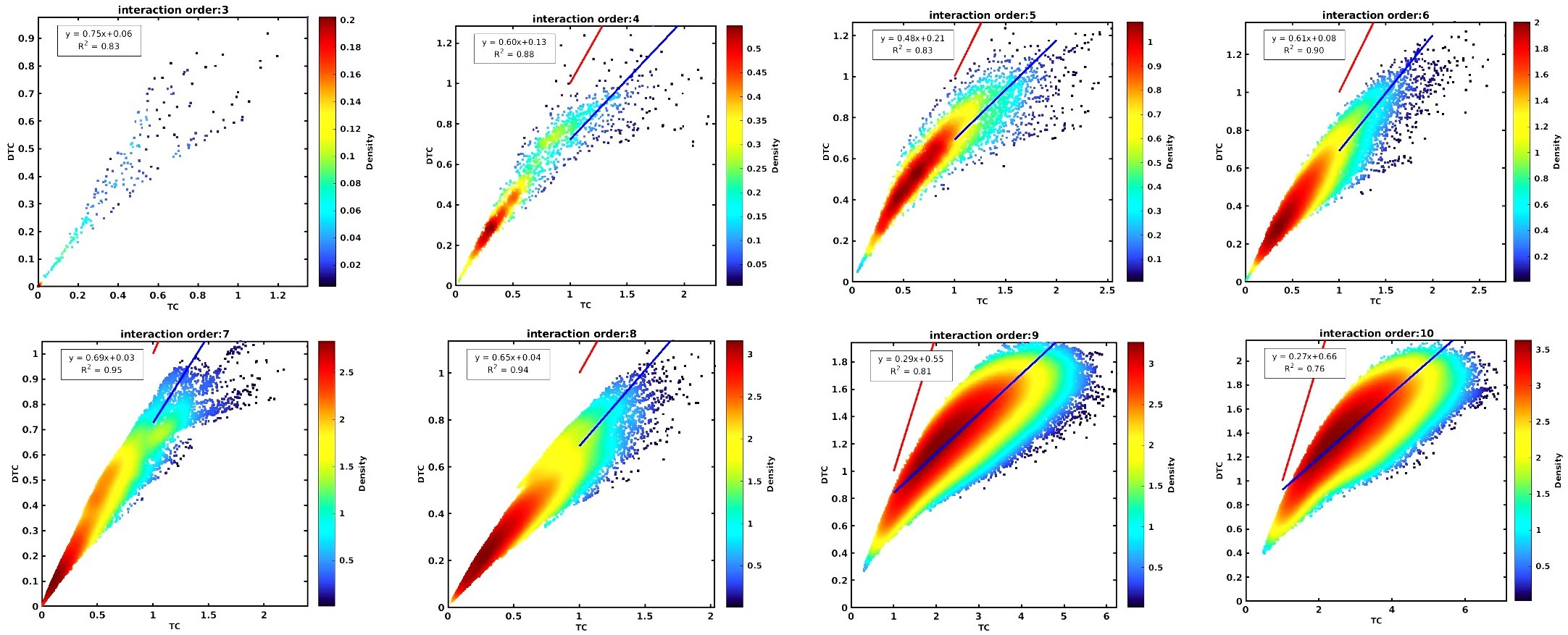}
    \caption{High-order interactions (HOI) capture richer information and reveal distinct interaction features under resting-state and task-based conditions. (Upper panel) When the interaction order is set to two, TC is equivalent to DTC under both resting-state and task conditions, as both measures reduce to mutual information in this case. (Middle panel) For the resting-state data, increasing the interaction order (from 3 to 10) led to the capture of progressively greater amounts of interaction information, suggesting the involvement of more complex and distributed brain interactions. (Lower panel) A similar pattern was observed in the task-based fMRI data. However, with higher interaction orders (9 and 10), the strength of interactions increased more prominently, indicating that completing the task engages stronger and more coordinated high-order interactions. Notably, this phenomenon was more evident at higher orders compared to low-order interactions.}
    \label{fig:45}
\end{figure*}

A clear order-dependent pattern was observed across both resting-state and task-based fMRI conditions (\textbf{\textit{Fig.\ref{fig:45}}}). At interaction order two, TC and DTC produced identical values across both conditions, reflecting their equivalence to mutual information at this level. This confirms that, at low order, both measures capture only pairwise statistical dependencies without distinguishing higher-order effects.

As the interaction order increased from 3 to 10 in the resting-state condition, the amount of captured interaction information increased progressively. This monotonic rise indicates that higher-order dependencies contribute additional information beyond pairwise interactions, reflecting the presence of distributed and multi-regional coordination in spontaneous brain activity. The steady growth across orders suggests that resting-state dynamics are supported by structured, large-scale interaction patterns that become increasingly apparent when higher-order relationships are considered.

A similar increasing trend was observed in the task-based condition; however, the magnitude of interaction strength became notably more pronounced at higher orders, particularly at orders 9 and 10. This amplification at high orders indicates that task engagement recruits more strongly coordinated multi-region interactions compared to rest. Importantly, the distinction between resting-state and task conditions became more evident at higher orders, whereas low-order interactions showed relatively limited differentiation. These findings demonstrate that high-order analysis provides additional discriminatory power and reveals interaction structures that are not captured by low-order measures alone.

\begin{figure*}
    \centering
    \includegraphics[width=0.84\textwidth, height=11.2cm]{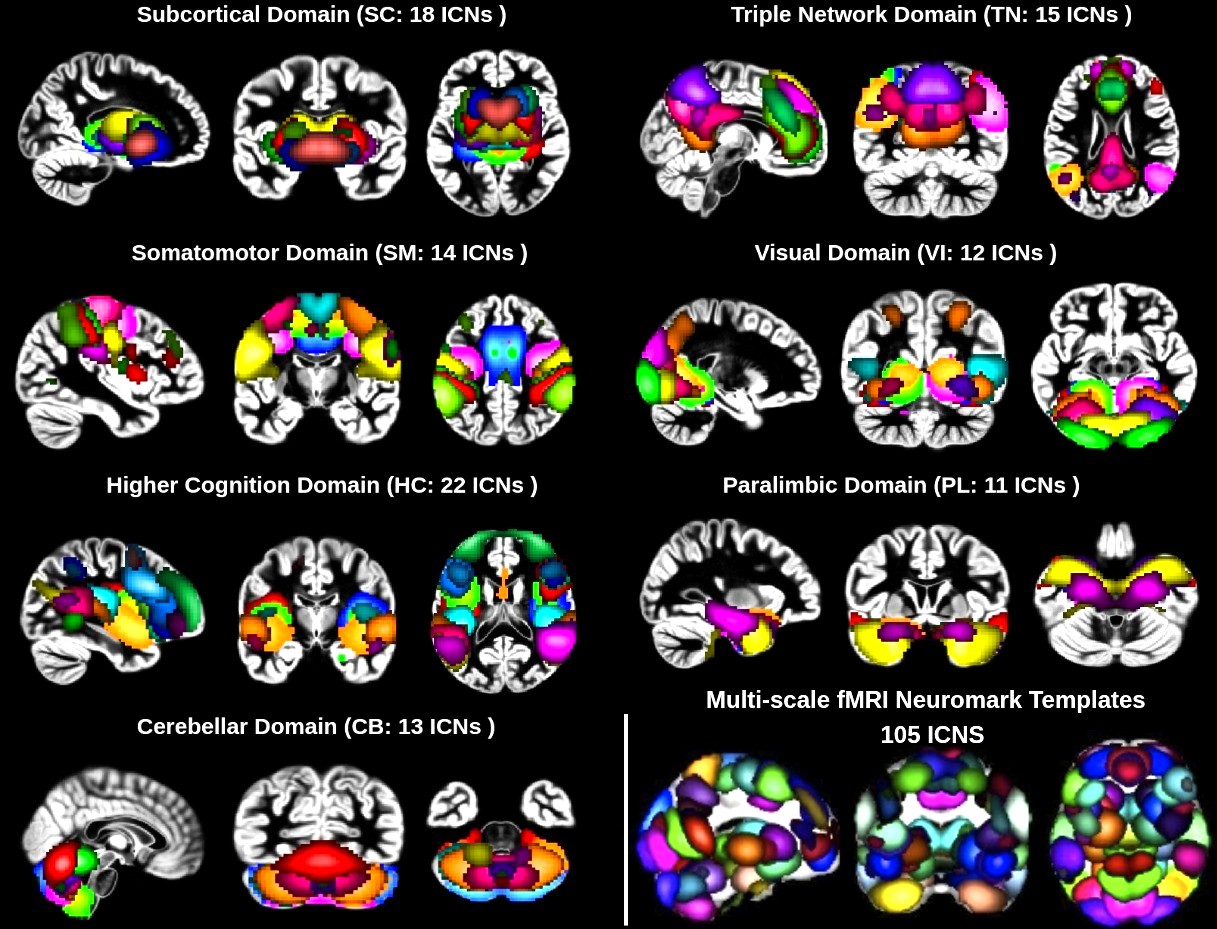}
    \rule{\textwidth}{0.5mm}
    \includegraphics[width=\textwidth, height=11.5cm]{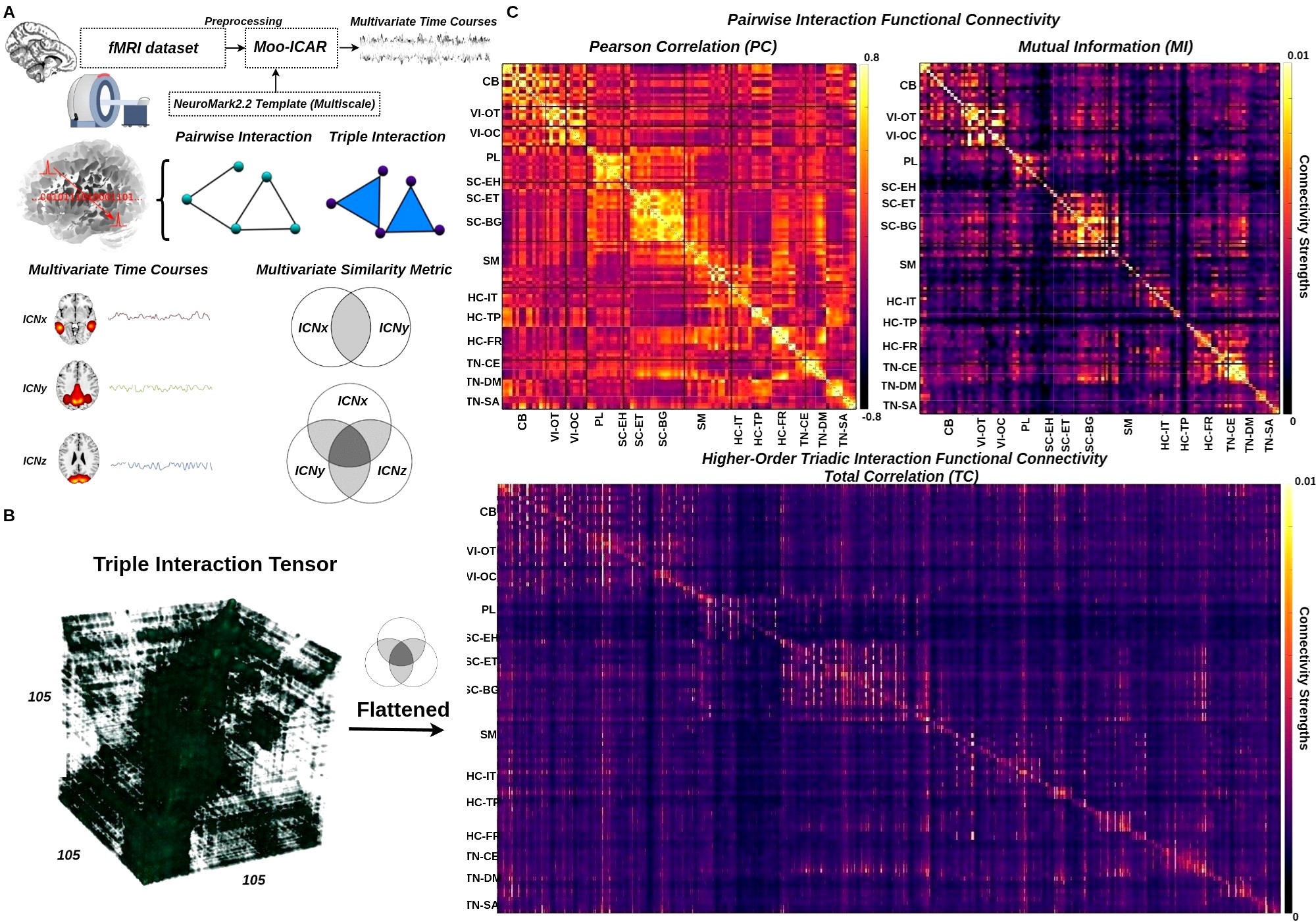}
    \label{fig:12}
\end{figure*}
\begin{figure*}
    \addtocounter{figure}{0}
    \caption{Functional connectomics of the human brain. (Upper panel) Multi-scale fMRI neuromark template~\cite{Iraji2022CanonicalAR}. The template includes 105 intrinsic connectivity networks (ICNs), which are organized into seven major brain domains: subcortical (SC), triple network (TN), somatomotor (SM), visual (VI), higher cognitive (HC), paralimbic (PL), and cerebellar (CB). (Lower panel) Large-scale resting-state networks were identified by applying constrained ICA to preprocessed fMRI data, resulting in 105 independent network components along with their corresponding time courses (\textbf{A}). High-order triadic functional connectivity was then constructed using total correlation, represented as a 3D tensor of size $105 \times 105 \times 105$. To enable comparison with pairwise connectivity, the triadic tensor can be flattened into a 2D matrix, which is also ordered according to network community (\textbf{B}). Pairwise functional network connectivity was computed using Pearson correlation and mutual information, and the resulting matrices were ordered according to network community (\textbf{C}).}
    \label{fig:12}
\end{figure*}

\subsection*{Large-Scale Functional Connectivity Networks}
To estimate intrinsic connectivity networks (ICNs) from our resting-state data, we used the multi-scale \textit{NeuroMark\_fMRI\_2.2} template~\cite{Iraji2022CanonicalAR}. This template, derived from over 100,000 subjects, contains 105 networks across multiple spatial resolutions. The ICNs were generated from more than 20 studies using a group multi-scale ICA approach with eight model orders. Incorporating multiple scales allows us to model a diverse range of intrinsic connectivity networks.

The 105 ICNs are grouped into seven major functional domains (\textbf{\textit{Fig.\ref{fig:12}}}): visual (VI, 12 ICNs; occipitotemporal (OT) and occipital (OC) subdomains), cerebellar (CB, 13), subcortical (SC, 18; extended hippocampal (EH), extended thalamic (ET), and basal ganglia (BG) subdomains), sensorimotor (SM, 14), higher cognition (HC, 22; insular-temporal (IT), temporoparietal (TP), and frontal (FR) subdomains), triple network (TN, 15; central executive (CE), default mode (DM), and salience (SA) subdomains), and paralimbic (PL, 11).

\subsection*{High-order Triadic Functional Connectivity}
To characterize high-order interactions (\textbf{\textit{Fig.\ref{fig:12}A}}), we computed triadic functional connectivity via a matrix-based entropy function, providing an estimate of total correlation, which quantifies statistical dependencies among all triplets of brain regions. For \(\vect{k} = 3\), the number of possible interactions increases dramatically, with a total of $105^{3} = 1,157,625$ interactions, corresponding to $\binom{105}{3} = 187,460$ unique triads. The triadic connectivity is represented as a three-dimensional tensor of size $105^{3}$. For comparison with pairwise functional connectivity, the tensor can be flattened into a two-dimensional matrix while preserving network community order (\textbf{\textit{Fig.\ref{fig:12}B}}). Triadic connectivity captures interaction patterns beyond what pairwise correlations reveal. By incorporating these high-order dependencies, it provides a more comprehensive framework for examining the complex, multidimensional organization of functional connectivity across multiscale human brain networks.

\subsection*{Pairwise Functional Connectivity}
Pairwise and triadic connectivity exhibit complementary organizational features. While pairwise functional network connectivity (FNC) highlights direct correlations between region pairs (\textbf{\textit{Fig.\ref{fig:12}C}}), triadic FC identifies multivariate dependencies, uncovering high-order network structures. These structures provide insight into distributed information integration across networks and offer a richer characterization of the functional architecture of the brain.

The combinatorial increase in interactions is substantial: compared to the $\binom{105}{2}=5,460$ unique pairwise connections, considering all possible triple interactions yields $1,157,625$ interactions, an increase by a factor of approximately $212.0$. Similarly, the number of unique triads, $187,460$, represents a $\sim 34.3$-fold increase relative to unique pairwise connections. Despite the inherent combinatorial complexity, the triadic framework uncovers organized clustering structures, indicating that high-order triadic connectivity reflects biologically meaningful network interactions.

\begin{figure*}[!ht]
    \centering
    \includegraphics[width=\textwidth, height=9cm]{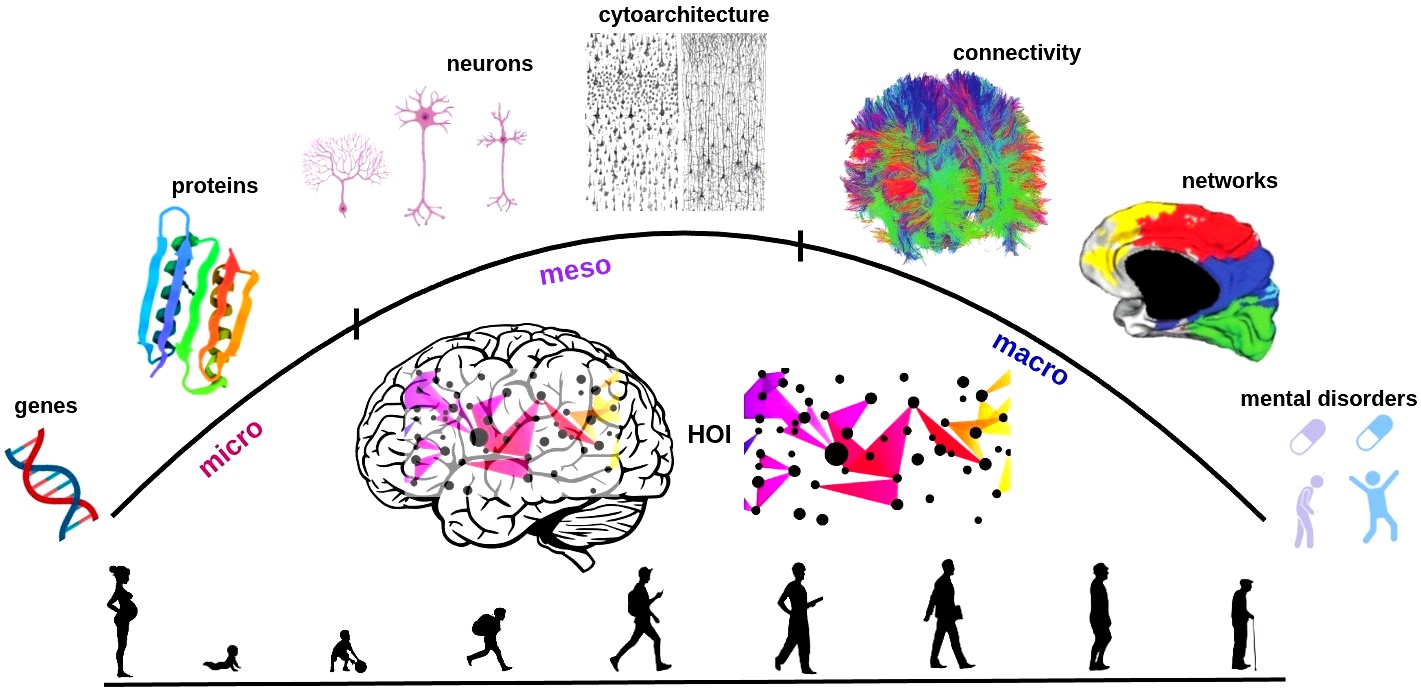}
    \caption{High-order interactions across the multiscale organization of the human brain over the lifespan. Selected examples illustrate microscale, mesoscale, and macroscale brain organization and (dys)function, arranged from left to right: genes, proteins, neurons, cytoarchitecture, connectivity, networks, and mental disorders.}
    \label{fig:3}
\end{figure*}

\section*{Challenges and Opportunities}
In summary, high-order triadic connectivity offers a promising framework for studying cooperative interactions among multiple brain regions. By capturing dependencies beyond pairwise correlations, this approach provides a closer approximation of the brain’s information-processing mechanisms and opens the way to move beyond traditional functional connectivity analyses. However, some challenges remain, which we address based on our experiments as follows.

\textbf{Biological Interpretation.} Triadic functional connectivity captures statistical dependencies among triplets of brain networks, revealing interactions that pairwise metrics cannot detect. Biologically, these triads can correspond to coordinated neural ensembles, convergent input to a hub, or distributed circuits that support high-order cognitive functions. Unlike pairwise correlations, which may conflate indirect interactions or shared variance, triadic measures can distinguish direct multivariate dependencies, providing insight into how the brain integrates information across multiple regions simultaneously. 

\textbf{Interaction Order.} The interaction order defines the number of networks considered simultaneously in the functional connectivity model. Pairwise connectivity corresponds to second-order interactions, whereas triadic connectivity is third-order, capturing dependencies that emerge only when three regions interact jointly. Extending to higher orders enables the identification of combinatorial networks and emergent structures invisible to lower-order analyses. However, high-order interactions also raise questions about interpretability and biological relevance. Additionally, the brain may mix or switch the number of interactions in different states.

\textbf{High-Order Complexity.} Estimating triadic interactions across a whole-brain connectome scales combinatorially with the number of regions (networks), which presents significant computational challenges. For a network with $N$ nodes, the number of unique triads is given by the binomial coefficient
\[
\binom{N}{3} = \frac{N!}{3!(N-3)!},
\]
which grows rapidly with increasing $N$. Consequently, computing triadic connectivity exhaustively in large-scale networks, and extending this analysis beyond triads, is computationally infeasible, increases effect size, and is challenging to visualize. At the same time, this raises an important question: how does the brain achieve fast and efficient communication despite such high-order complexity?

To address this, several strategies can be employed. Sparse estimation techniques focus on a subset of statistically significant interactions, reducing the number of triads considered. Tensor decomposition approaches allow the representation of high-order interactions in a compressed form, while graph-theoretical motif analysis can approximate triadic patterns without enumerating all possible triplets. These approaches balance computational tractability with the retention of biologically meaningful information. 

\textbf{Extension to O-/S-information, and Partial Information Decomposition.} While TC provides a global measure of collective dependencies, it does not distinguish between redundant, unique, or synergistic contributions. To address this, it is necessary to extend the analysis using the O-information and S-information~\cite{rosas2019quantifying}, which distinguish between redundant and synergistic contributions in high-order interactions. The O-information quantifies the balance between redundancy and synergy in a system, while the S-information highlights synergistic contributions. Together, these metrics enable a more nuanced characterization of multivariate dependencies, particularly in complex networks such as the brain.

Finally, the framework of Partial Information Decomposition (PID)~\cite{Williams2010NonnegativeDO} provides a rigorous formalism to separate multivariate information into unique, redundant, and synergistic components for subsets of variables. By combining TC, DTC, and PID-based measures, we can systematically quantify both pairwise and high-order interactions in functional brain networks, offering a comprehensive description of information flow and collective dynamics.

\textbf{Dynamics.} Triadic interactions are inherently dynamic, reflecting fluctuations in multivariate dependencies over time. Dynamic triadic connectivity captures transient cooperation, revealing how information is routed through networks during resting-state fluctuations or pathological conditions. Temporal analysis of these interactions can uncover state-dependent network reorganization, detect early disruptions in psychiatric disorders, and inform models of neural computation that depend on coordination across multiple regions rather than on pairwise interactions.

\textbf{Applications in Cognitive Science, Mental Health, and Beyond.} Triadic functional connectivity provides a powerful framework for understanding complex brain function and dysfunction across different levels, from micro to meso to macro, over the lifespan (\textbf{\textit{Fig.\ref{fig:3}}}). In cognitive neuroscience, it reveals how multiple regions, as well as genes, proteins, and neurons, jointly encode or integrate information, going beyond pairwise correlations. In mental health, it can uncover subtle network disruptions across molecular, cellular, and neural levels that are linked to psychiatric disorders, enhancing the identification of potential biomarkers. Beyond neuroscience, this encompasses high-order network modeling of complex systems and signal analysis across any modality, where multivariate interactions are both crucial and meaningful.

\begin{acknowledgments}
This work was supported by NSF grant 2112455, and NIH grants R01MH123610 and R01MH119251.
\end{acknowledgments}

\section*{Author contributions}
Q. Li., VD. Calhoun.: Conceptualization, Investigation, Software, Writing – Review \& editing. Q. Li., M. Seraji., YP. Wang., GD. Pearlson., VD. Calhoun.: Writing – Review \& editing. VD. Calhoun.: Funding Acquisition.

\section*{Declaration of interests}
The authors declare that they have no known competing financial interests or personal relationships that could have appeared to influence the work reported in this paper.

\bibliography{pnas-sample,apssamp}
\end{document}